\documentclass[prb,aps,showpacs,twocolumn,floatfix] {revtex4}
\usepackage{graphicx,color}
\usepackage{dcolumn}
\usepackage{bm}

\newcommand{\quadr}[1]{ \left[ #1 \right] }
\newcommand{\gr} [1]{\textbf #1}
\newcommand{\kg}{ \textbf{k} }
\newcommand{\tonda}[1]{ \left( #1 \right) }
\newcommand{\graf}[1]{ \left\{ #1 \right\} }

\begin{document}

\title{
\bf{The Ising phase in the $J_1-J_2$ Heisenberg Model}
}

\author{
Valeria Lante
and Alberto Parola
}

\affiliation {
 Dipartimento di Fisica e Matematica, Universit\'a dell'Insubria, 
Via Valleggio 11
Como, Italy }

\date{\today}

\begin{abstract}
The two dimensional Heisenberg antiferromagnet on the square lattice with 
nearest ($J_1$) and next-nearest ($J_2$) neighbor couplings is investigated 
in the strong frustration regime ($J_2/J_1>1/2$). 
A new effective field theory describing the long wavelength physics of the model
is derived from the quantum hamiltonian. The structure of the resulting 
non linear sigma model allows to recover the known spin wave results in the 
collinear regime, supports the presence of an
Ising phase transition at finite temperature and suggests the possible occurrence
of a non-magnetic ground state breaking rotational symmetry.
By means of Lanczos diagonalizations we investigate the spin system 
at $T=0$, focusing our attention on the region where the collinear order parameter
is strongly suppressed by quantum fluctuations and a transition to a non-magnetic state occurs. 
Correlation functions display a remarkable size independence and allow to identify the 
transition between the magnetic and non-magnetic region of the phase diagram.  
The numerical results support the presence of a non-magnetic phase with orientational ordering.
\end{abstract}

\maketitle

\section{Introduction}

Frustrated low-dimensional spin systems are still extensively investigated since 
they show rich phase diagrams and exhibit unusual quantum phases.
A typical example is the square lattice $J_1-J_2$ model: a Heisenberg 
antiferromagnet with 
competing couplings ($J_1,J_2>0$) between nearest neighbor ($<\, >$) 
and next-nearest neighbor ($\ll\, \gg$) spins:
\begin{equation} \label{frustrato0}  
\mathcal{H}=\emph{J}_1\sum_{<i,j>}\hat{\gr{S}}_i \cdot \hat{\gr{S}}_j+{J}_2
\sum_{\ll i,k \gg}\hat{\gr{S}}_i \cdot \hat{\gr{S}}_k 
\end{equation}
where $\hat{\gr{S}}_i$ are spin operators. By varying the frustration ratio 
$\alpha=J_2/J_1$, quantum phase transitions between magnetically ordered and 
disordered phases can take place at $T=0$. The interest in the 2D frustrated 
antiferromagnets has deep theoretical motivations in the characterization of 
disordered spin liquids (or Bose liquids). Moreover, several magnetic
materials have been synthesized nowadays where frustration plays a dominant role 
\cite{varie}. In particular, the specific interest on the frustrated Heisenberg model on
a square lattice raised with the discovery of vanadate compounds, whose magnetic 
behavior is likely to be described by the $J_1-J_2$ hamiltonian \cite{Melzi}.   

In the classical limit ($S\to\infty$) at weak frustration,
the ground state (GS) has conventional N\'eel
order with magnetic wave vector $\gr{Q}=(\pi,\pi)$ for $\alpha<0.5$. Above this threshold,
the two sublattices are antiferromagnetically 
ordered but remain free to rotate with respect to each other and the 
GS manifold has an $O(3)\times O(3)$ degeneracy, larger than expected on the
basis of the $O(3)$ symmetry of the hamiltonian. Weak quantum 
fluctuations, via the  \emph{order by disorder} mechanism, can be included by use of
spin wave theory, and are shown to select a
collinear ordered state with magnetic wave vector $\gr{Q}=(\pi,0)$ or $(0,\pi)$, 
reducing the GS degeneracy to $O(3)\times Z_2$. 
In fact, the collinear state breaks both the $O(3)$ spin rotational invariance
of the Heisenberg hamiltonian and the $\pi/2$ rotational symmetry of the square lattice $(Z_2)$.
In the $S\to\infty$ limit, according to spin wave theory \cite{cd}, the critical coupling 
$\alpha_c=1/2$ marks the first order transition between the collinear and the N\'eel phase. 
However, when quantum fluctuations are taken into account beyond perturbation theory, 
(at least) one intermediate phase is expected to separate the two 
magnetically ordered phases \cite{varie}.

Nevertheless, on the basis of a very recent perturbative numerical 
renormalization group analysis no evidence for an intermediate 
phase has been found and it
 has been proposed that a direct and unexpected second order phase 
transition may occur at the classical critical point\cite{MOU}.\\
The aim of this work is to investigate the quantum phase transition occurring 
at $T=0$ as the frustration ratio $\alpha$ is decreased, i.e. when the collinear 
order is suppressed by quantum fluctuations. 
To this purpose, the effective long-wavelength action 
of the two dimensional quantum model in the regime of strong frustration
 is mapped into a non-linear sigma model (NLSM) in 2+1 dimension.
This is the first time that the mapping, originally proposed by Haldane for 
a Heisenberg chain \cite{H}, is properly generalized to the strongly
 frustrated $J_1-J_2$ model.
On the basis of the symmetries of the NLSM action, an Ising phase, breaking 
the $\frac{\pi}{2}$ rotational symmetry of the square lattice and preserving 
the $O(3)$ spin rotational invariance, may exist both at low and zero temperature.
While the stability of this phase at finite temperature has been previously 
investigated \cite{CCL,singh, caprio1,caprio2}, it has never been explicitly 
examined if such an Ising state is stable
at zero temperature for some values of the frustration ratio lower than 
the one corresponding to the onset of the collinear order.
Lanczos diagonalizations support this scenario via a careful examination 
of both the excitation spectrum and the correlation functions:
A non-magnetic valence bond nematic phase with orientational ordering
is the most favorable candidate as a ground state in a portion of the phase diagram.

\section{Non linear Sigma Model for the collinear phase}

Among the various theoretical approaches adopted for the $J_1-J_2$ 
model in the regime of weak frustration, the NLSM 
method is particularly suitable for the study of the phase transition 
between a magnetically ordered state and a disordered phase.
The $2D$ frustrated Heisenberg antiferromagnet 
with $\alpha<\alpha_c$ is mapped to a $O(3)$ NLSM in $D=2+1$ 
dimension \cite{Chak,takano}, 
which indeed shows a second order (quantum) phase transition to a 
non-magnetic state at $T=0$. When frustration is strong and the GS is collinear, 
the mapping to a $3D$ classical model is still possible, 
but the effective long-wavelength 
action is no longer a conventional $O(3)$ NLSM. 

Here we generalize the original mapping proposed by Haldane for the microscopic derivation 
of the long-wavelength, low-energy effective theory of one-dimensional 
quantum antiferromagnets in
the N\'eel phase \cite{H}, to the $J_1-J_2$ 
model in the strong frustration regime, where collinear order is expected. 
By using the Trotter formula and a coherent state basis in the spin Hilbert 
space, 
the partition function of the system is written in a path integral
 representation as:
\begin{equation} \label{partizionefinale} 
\mathcal{Z}=\int \mathcal{D}\hat\Omega\,\exp{\tonda{-\mathcal{S}[\hat\Omega]}} 
\end{equation}
where $\hat\Omega_i(t)$ is a classical $O(3)$ vector field defined 
on each lattice site,
normalized to $\hat\Omega^2_i(t)=1$ and $\mathcal{S}$ is the action at a temperature $\beta^{-1}$
\begin{equation}\label{azioneS} 
\mathcal{S} [\hat\Omega\tonda{\tau}]=-iS\sum_i\omega[\hat{\Omega}_i(\tau)]+
\int_0 ^{\beta}d\tau\gr{H}[\hat{\Omega}(\tau)]
\end{equation}
The first (purely imaginary) term is the Berry phase \cite{H} and 
$\gr{H}[\hat{\Omega}]$ is the expectation value of the 
hamiltonian operator on the coherent states basis:
\begin{equation}\label{hantiffrustr}
\gr{H}[\hat{\Omega}]=\emph{S}\,^2\emph{J}_1\sum_{<i,j>}\hat{\Omega}_i 
\cdot \hat{\Omega}_j+\emph{S}\,^2{J}_2\sum_{\ll i,k \gg}\hat{\Omega}_i \cdot \hat{\Omega}_k
\end{equation}
In the regime of strong frustration it is convenient to separate the lattice in
the two sublattices which will be labeled $+$ and $-$ respectively.
Following Haldane \cite{H}, we split the spatially oscillating 
spin state $\hat\Omega_i$ on the sublattice $+$ ($-$) as the sum 
of two orthogonal smooth vector 
fields, describing the local N\'eel order $\hat\gr{n}_{+}$ ($\hat\gr{n}_{-}$) 
and the transverse fluctuations $\gr{L}_{+}$ ($\gr{L}_{-}$),
 satisfying the constraints
$\hat\gr{n}_{+}^2=1$ ($\hat\gr{n}_{-}^2=1$)  
and $\hat\gr{n}_{+}\cdot\gr{L}_{+}=0$ ($\hat\gr{n}_{-}\cdot\gr{L}_{-}=0$). 
{In order to  carry out the splitting between the uniform and staggered
fluctuations keeping the right number of independent variables,
we proceed by partitioning the lattice in plaquettes as shown in Fig.
\ref{lattice}. Then we define the N\'eel fields 
($\hat{\gr{n}}_{+}$, $\hat{\gr{n}}_{-}$) and the associated fluctuations 
($\gr{L}_+$, $\gr{L}_-$) in the center of each plaquette as follows:
\begin{equation}\label{omegaA}
\hat{\Omega}(x+\eta\frac{a}{2},y+\eta\frac{a}{2})=
\eta\,\hat{\gr{n}}_{+}(\gr{r})\sqrt{1-\Big\vert\frac{\gr{L}_{+}(\gr{r})}{S}{\Big\vert}^2}+
\frac{\gr{L}_{+}(\gr{r})}{S}
\end{equation} 
\begin{equation}\label{omegaB}
\hat{\Omega}(x+\eta\frac{a}{2},y-\eta\frac{a}{2})=
\eta\,\hat{\gr{n}}_{-}(\gr{r})\sqrt{1-\Big\vert\frac{\gr{L}_{-}(\gr{r})}{S}{\Big\vert}^2}+
\frac{\gr{L}_{-}(\gr{r})}{S}
\end{equation} 
where $\eta=\pm 1$, $a$ is the spacing of the original lattice and 
$\gr{r}=\tonda{x,y}$ is the coordinate of the plaquette centers
which define a square superlattice of spacing $2a$.
\begin{figure} \label{fig1}
\includegraphics[width=0.4\textwidth]{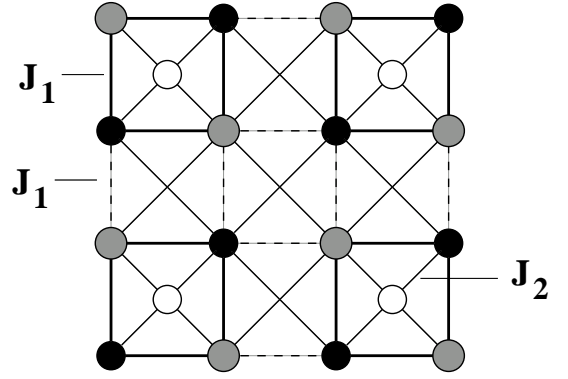}
\vspace{-3mm}
\caption{\label{lattice}
Lattice of the $J_1-J_2$ model. The lattice spacing is $a$. Grey and black 
circles denote $+$ and $-$ sublattice sites respectively, while white circles 
the center of each plaquette, where the fields $\hat{\gr{n}}_{+}$, 
$\hat{\gr{n}}_{-}$ and 
$\mathrm{L}_{+}$, $\mathrm{L}_{-}$ are defined.
}
\vspace{-5mm}
\end{figure}
In the continuum limit, to second order in space and time derivatives
and keeping the lowest order in $1/S$, the partition function is written as
\begin{equation}
\mathcal{Z}=\int \mathcal{D}\hat{\gr{n}}_\pm\,\mathcal{D}{\gr{L}}_\pm\,\delta\,{(\hat{\gr{n}}_+\cdot\gr{L}_+)}
\delta\,{(\hat{\gr{n}}_-\cdot\gr{L}_-)}\,e^{-S}
\label{zed}
\end{equation}
where 
$$
S ={\int\, d^2\gr{x}\int_0^{\beta} d\tau\mathcal{L}} 
$$
and the lagrangian is 
\begin{equation}
\mathcal{L} = 
\mathrm{L}^T\mathrm{A}\mathrm{L}-\mathrm{B}^T\mathrm{L}+\mathrm{K}_1+\mathrm{K}_2
\label{formal}
\end{equation}
$\mathcal{D}\hat{\gr{n}}_\pm$ and $\mathcal{D}{\gr{L}}_\pm$ mean 
integration over the N\'eel fields and the fluctuations in both the 
sublattices and the six component array
 $\mathrm{L}^T=(\mathrm{\gr{L}}_+,\mathrm{\gr{L}}_-)$ gathers the (three)
components of the two fluctuation fields $\mathrm{L}_+^\alpha$ 
and $\mathrm{L}_-^\alpha$. 
The scalars $\mathrm{K}$ are defined as
\begin{eqnarray}
\mathrm{K}_1&=&-{S^2\emph{J}_1}\tonda{\partial_y\hat{\gr{n}}_+\cdot\partial_y\,\hat{\gr{n}}_-
-\partial_x\,\hat{\gr{n}}_+\cdot\partial_x\,\hat{\gr{n}}_-} \nonumber\\
\mathrm{K}_2&=&{S^2\emph{J}_2}\sum_{k=\pm}
\quadr{(\nabla{\hat{\gr{n}}_k})^2+k\,
\partial_x\hat{\gr{n}}_k \cdot\partial_y\hat{\gr{n}}_k}
\end{eqnarray}
while the $6\times 6$ matrix $\mathrm{A}$ is written in block form as
\begin{eqnarray}\label{matriceA}
\mathrm{A}\,=\,\frac{1}{a^2}\left(\begin{array}{ccc}
{2\emph{J}_2} & {\emph{J}_1} \\
{\emph{J}_1} & {2\emph{J}_2}
\end{array} \right)
\end{eqnarray}
and the array $\mathrm{B}^T=\tonda{\mathrm{\gr{B}}_+,\mathrm{\gr{B}}_-}$ is given by:
\begin{eqnarray}
\mathrm{\gr{B}}_+\,&=&\,\frac{-i}{2a^2}\tonda{\hat{\gr{n}}_+\times\partial_{\tau}
{\hat{\gr{n}}_+}}+\frac{SJ_1}{a}\vec{\gamma}_-+\frac{2SJ_2}{a}\vec{\gamma}_+\nonumber\\
\mathrm{\gr{B}}_-\,&=&\,\frac{-i}{2a^2}\tonda{\hat{\gr{n}}_-\times\partial_{\tau}
{\hat{\gr{n}}_-}}+\frac{SJ_1}{a}\vec{\gamma}_++\frac{2SJ_2}{a}\vec{\gamma}_-\nonumber
\end{eqnarray}
with
$$
\vec{\gamma}_+\,=\,\partial_x \hat{\gr{n}}_+ + \partial_y \hat{\gr{n}}_+  \qquad
\vec{\gamma}_-\,=\,\partial_x \hat{\gr{n}}_- - \partial_y \hat{\gr{n}}_-
$$
To this order in derivatives the Berry phase contributions 
$i\omega[\hat{\gr{n}}_+(\tau)]$ and $i\omega[\hat{\gr{n}}_-(\tau)]$ 
identically vanish \cite{note}. Performing the gaussian integration of the 
partition function with respect to fluctuations needs some care due to the constraints 
$\hat{\gr{n}}_+\cdot{\gr{L}}_+=0$ and $\hat{\gr{n}}_-\cdot{\gr{L}}_-=0$
 which limit the integration to the
transverse components of the spin fluctuations ${\gr{L}}^{\perp}_+$ 
and ${\gr{L}}^{\perp}_-$. 
In order to release the constraint, 
we first multiply the partition function by a constant factor $F$ (which does not
affect the physical properties of the model) written as 
a gaussian integral over two auxiliary scalar fields $v_+$ and $v_-$: 
$$
F=\int \mathcal{D} v_\pm \,e^{-\int\, d^2\gr{x}\int_0^\beta d\tau \, ({\mathrm{L}}^{\parallel})^T A 
{\mathrm{L}}^{\parallel}}
$$
where ${\gr{L}}^{\parallel}_+=v_+ \hat{\gr{n}}_+$ 
(${\gr{L}}^{\parallel}_-=v_- \hat{\gr{n}}_-$). By defining the vector 
${\gr{L}}_+ = {\gr{L}}^{\perp}_+ + {\gr{L}}^{\parallel}_+$ 
 (${\gr{L}}_- = {\gr{L}}^{\perp}_- + {\gr{L}}^{\parallel}_-$) ,
the integral over fluctuations $\mathcal{I}$ in Eq. (\ref{zed}) can be written as:
$$
\mathcal{I}=\int \mathcal{D}{\gr{L}}_\pm\,
e^{ -\int\, d^2\gr{x}\int_0^{\beta} d\tau 
\left [({\mathrm{L}}^\perp)^T {\mathrm{A}}{\mathrm{L}}^\perp
+ ({\mathrm{L}}^{\parallel})^T A {\mathrm{L}}^{\parallel}
- {\mathrm{B}}^T{\mathrm{L}}^\perp  \right ]}
$$
where ${\gr{L}}^{\perp}_+$ is explicitly given by ${\gr{L}}^{\perp}_+={\gr{L}}_+
-\tonda{\gr{L}_+ \cdot{\hat{\gr{n}}}_+} {\hat{\gr{n}}}_+$ ( ${\gr{L}}^{\perp}_-={\gr{L}}_-
-\tonda{\gr{L}_- \cdot{\hat{\gr{n}}}_-} {\hat{\gr{n}}}_-$). As a result, the partition
function is written as an unconstrained integral over the fields ${\gr{L}}_\pm$ and ${\hat{\gr{n}}}_\pm$.
The lagrangian maintains the same formal structure (\ref{formal}) with a modified
matrix $\mathrm{A}$ and array $\mathrm{B}$:
\begin{eqnarray}
\tilde{\mathrm{A}}_{ik}^{\alpha \beta}&=&\mathrm{A}_{ik}
\tonda{\delta^{\alpha \beta}-n_i^\alpha n_i^\beta-n_k^\alpha n_k^\beta+ 2(n_i 
\cdot n_k)\, n_i^\alpha n_k^\beta} \nonumber \\
\tilde{\mathrm{B}}_i^\alpha&=&{\mathrm{{B}}}_i^\alpha-\tonda{{\mathrm{\gr{B}}}_i\cdot \hat{\gr{n}}_i} 
\hat{n}_i^\alpha 
\end{eqnarray}
where Latin indices $i,k=\pm$ identify the sublattice and Greek 
superscripts run over the three spin components.
In practice, the gaussian integration has been performed by diagonalizing 
the matrix $\tilde{\mathrm{A}}$ on its eigenvector basis 
$\graf{u_i ^{\alpha} \tonda{l}}_l$ with $l=1, \dots ,6$ so that the effective lagrangian density is 
expressed in terms of the eigenvalues $\graf{\lambda_l}_l$ 
\begin{equation}
\mathcal{L} = \mathrm{K_1}  + \mathrm{K_2} - \frac{1}{4} \sum_{l} \lambda _l ^{-1} b_l^2 
\qquad b_l = \sum_{i,\alpha}\tilde{\mathrm{B}}_i^{\alpha} u_i ^{\alpha}\tonda{l}
\end{equation}
After tedious but straightforward calculations, the resulting effective lagrangian is written as
\begin{equation}
\mathcal{L}[\hat{\gr{n}}_+,\hat{\gr{n}}_-]=\mathcal{L}_S+\mathcal{L}_A+\mathcal{L}_\times 
\label{final}
\end{equation}
with
\begin{eqnarray}
\mathcal{L}_S &=&\sum_{i=\pm}\left \{\frac{S^2\emph{J}_2}{2}\,\vert\nabla\hat{\gr{n}}_i
\vert^2 +\frac{\emph{J}_2}{8\,a^2\,D^*}
\,\tonda{\partial_\tau\hat{\gr{n}}_i}^2 \right \}\nonumber \\
\mathcal{L}_A &=& -\frac{S^2\emph{J}_1}{2}\tonda{\partial_y\hat{\gr{n}}_+\cdot\partial_y
\hat{\gr{n}}_--\partial_x\hat{\gr{n}}_+\cdot\partial_x\hat{\gr{n}}_-} \nonumber \\
\mathcal{L}_\times&=& 
\frac{\emph{J}_1\emph{J}_2\,^2}{2a^2D}{\tonda{\hat{\gr{n}}_+\cdot
\partial_\tau\hat{\gr{n}}_-}\tonda{\partial_\tau\hat{\gr{n}}_+\cdot\hat{\gr{n}}_-}}\,+ 
\nonumber\\
&+&\frac{\emph{J}_1\,^2\emph{J}_2}{8a^2D}{\quadr{\tonda{\hat{\gr{n}}_+\!
\cdot\partial_\tau\hat{\gr{n}}_-}^2+\tonda{\partial_\tau\hat{\gr{n}}_+\!\cdot\hat{\gr{n}}_-}^2}}
+ \nonumber \\
&-& \frac{\emph{J}_1}{8a^2D^*}\sigma\tonda{\partial_\tau\hat{\gr{n}}_+\cdot\partial_\tau\hat{\gr{n}}_-}
\nonumber 
\end{eqnarray}
Where $D^*={{4\emph{J}_2\,^2-\sigma^2\emph{J}_1\,^2}}$, $D=D^*(4\emph{J}_2\,^2-\emph{J}_1\,^2)$ 
and $\sigma = \hat{\gr{n}}_+\cdot\hat{\gr{n}}_-$ is the ``Ising order parameter".
The lagrangian density $\mathcal{L}$ preserves the $O(3)$ symmetry of the
microscopic hamiltonian (\ref{frustrato0}) but breaks the invariance of the model 
under $\pi /2$ lattice rotation due to the adopted coarse graining
procedure \cite{CCL}. However, the original rotational invariance 
reflects in the additional $Z_2$ symmetry of our lagrangian under the simultaneous 
action of $\pi /2$ rotation and inversion of one field: $(x,y) \to (-y,x)$ {\sl and}
$\hat{\gr{n}}_+ \to -\hat{\gr{n}}_+$. The global symmetry group of the resulting
field theory is therefore $O(3)\times Z_2$. 

In order to check the correctness of the effective lagrangian density $\mathcal{L}$, 
we compare the dispersion relation arising from a saddle point 
evaluation of the partition function to the known spin wave results. 
The saddle point configuration corresponds to the minimum of the action, i.e. to
a homogeneous and static configuration characterized by two independent unit
vectors $\hat{\gr{n}}_+^0 $ and $\hat{\gr{n}}_-^0 $ which describe one of the degenerate 
classical ground states. By expanding up to second order in fluctuations 
$\hat{\gr{n}}_+=\hat{\gr{n}}_+^0+\delta\hat{\gr{n}}_+$ 
($\hat{\gr{n}}_-=\hat{\gr{n}}_-^0+\delta\hat{\gr{n}}_-$) 
and performing the gaussian integration we get precisely 
the same result of spin wave theory in the long wavelength limit \cite{CCL,dagotto}: 
The low energy excitations are described by four branches labeled by $\lambda=\pm1$ and $\eta=\pm 1$
whose dispersions are given by 
\begin{eqnarray}
\omega^2(\kg)&=&4 S^2\{k_x^2\tonda{2\emph{J}_2+\eta\emph{J}_1}\tonda{2\emph{J}_2+\eta\emph{J}_1\cos{\theta}}+\nonumber\\&+& k_y^2\tonda{2\emph{J}_2+\lambda\emph{J}_1}\tonda{2\emph{J}_2-\lambda\emph{J}_1\cos{\theta}}\}
\label{swt}
\end{eqnarray}
where $\theta$ is the angle between 
the two N\'eel fields: $\cos\theta=\hat{\gr{n}}_+^0 \cdot \hat{\gr{n}}_-^0 $. 
Accordingly, perturbation theory on the 
effective action shows that the lowest free energy is attained when the
staggered magnetizations of the two sublattices $\hat{\gr{n}}_+^0 $ and 
$\hat{\gr{n}}_-^0 $ are either parallel or 
antiparallel: The collinear order is stabilized by quantum fluctuations.
This picture is believed to be correct in the limit of large frustration ratio $\alpha$.
By lowering $\alpha$, fluctuations are enhanced and a quantum phase transition is expected to occur
at zero temperature before reaching the (classical) limiting value $\alpha=0.5$. 
On the basis of the symmetries of the effective model, 
an intermediate regime characterized by vanishing
staggered magnetization $<\hat{\gr{n}}_+>\,=\,<\hat{\gr{n}}_->\,=\,0$ and possibly a finite Ising order 
parameter $<\sigma>=\hat{\gr{n}}_+\cdot\hat{\gr{n}}_-\,\ne 0$ may be present. 
This hypothetical state would break the 
rotational symmetry of the lattice preserving the $SU(2)$ spin symmetry. 
Due to the coarse graining carried out in the microscopic hamiltonian,
such a state may correspond either to a translationally invariant `` valence bond nematic"
phase where valence bonds display orientational ordering or to a valence 
bond crystal (VBC) which breaks the translational symmetry. 

A low energy, long wavelength effective field theory for 
collinear antiferromagnets was put forward in a seminal paper by 
Chandra, Coleman and Larkin \cite{CCL}. At variance with 
our approach, in that work quantum
fluctuations were integrated out by use of a perturbative spin-wave
approximation and the analysis was limited to the finite temperature domain.
In this way, CCL predicted a finite temperature Ising transition in
two dimensions: Even if at $T=0$ both sublattices display finite staggered magnetization in zero field, 
long-range N\'eel order disappears at any finite temperature, 
since thermal fluctuations are known to restore the continuous symmetries in two 
spatial dimensions \cite{mermin}. CCL suggested that the previously defined 
Ising order parameter $\sigma$ may preserve long-range order up to a non-zero critical temperature.
Recently the presence of such a finite temperature transition has been the subject of 
several investigations \cite{caprio1,caprio2,singh}.

The CCL approach, however, cannot be directly applied to the 
study of the zero temperature limit because the effects of quantum fluctuations are 
considered only within perturbation theory. In order to clarify the relationship between 
our effective field theory and the CCL approach, it is convenient to specialize
the lagrangian $\mathcal{L}[\hat{\gr{n}}_+,\hat{\gr{n}}_-]$ to a class of 
field configurations of the form 
$\hat{\gr{n}}(\gr{r},t)_+=\hat{\gr{n}}(\gr{r})_+^0+\delta\hat{\gr{n}}(\gr{r},t)_+$ 
($\hat{\gr{n}}(\gr{r},t)_-=\hat{\gr{n}}(\gr{r})_-^0+\delta\hat{\gr{n}}(\gr{r},t)_-$) 
characterized by weak fluctuations ($\delta\hat{\gr{n}}(\gr{r},t))$ on top of a
slowly varying {\sl time independent} (i.e. classical) configuration ($\hat{\gr{n}}(\gr{r})$).
Expanding $\mathcal{L}$ to second order in the fluctuations and performing the gaussian
integration, by use of the result (\ref{swt}), we recover the CCL result. 
This approach can be justified at large frustration ratios $\alpha$ (i.e. 
deep in the collinear regime) where quantum fluctuations are not able to 
severely affect the classical ground state. However, in order to analyze the 
quantum transition between collinear order and a disordered phase at $T=0$, 
it is necessary to take into account the effects of quantum fluctuations beyond 
the spin wave approximation, i.e. we have to study the full effective lagrangian
(\ref{final}).

\section{Lanczos Diagonalizations}
By means of Lanczos diagonalizations (LD) we try to clarify
if an Ising state can be stabilized when the long range collinear order 
fades away, as suggested by the field theoretical approach.

Lanczos diagonalizations have been performed on the $4 \times 4$ and 
$6 \times 6$ square clusters for spin $S=1/2$. By means of LD we obtain 
the energy spectrum, providing indications of possible 
changes in the nature of the GS which occur by increasing the frustration ratio \cite{Poil,ziman}.
In the collinear phase, due to the spatial symmetry breaking, four classes ({\sl towers})
of states with different spatial symmetries are expected to become degenerate: the lowest 
representative of these classes are an s-wave and a d-wave singlet at momentum 
$(0,0)$ and two triplets at momenta $(0,\pi)$ and $(\pi,0)$.
The low energy spectrum as a function of the frustration ratio is shown in Fig.\ref{gap}.
\begin{figure} \label{fig2}
\includegraphics[width=0.48\textwidth, height=0.42\textwidth ]{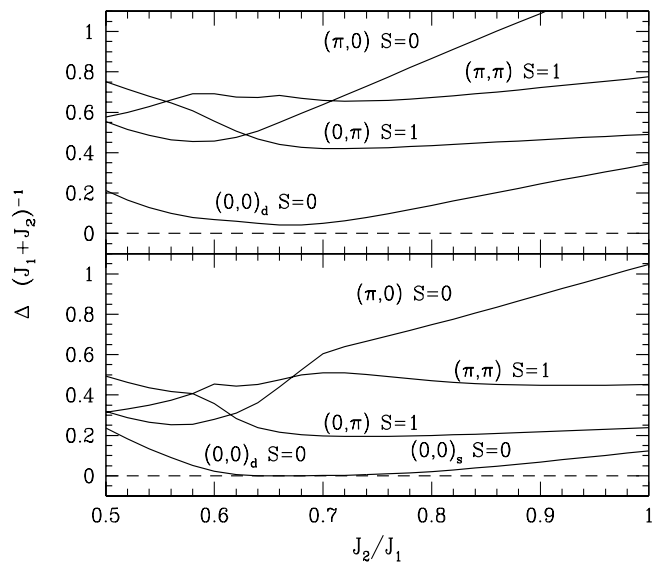}
\vspace{-4mm}
\caption{\label{gap}
Low energy states referenced to the GS as a function of the 
frustration ratio $\alpha=J_2/J_1$ in the $4\times 4$ (upper panel) and $6\times 6$ 
(lower panel) cluster.
}
\vspace{-5mm}
\end{figure}
While the behavior of the model for $\alpha\sim 0.5$ has been the subject of 
several investigations \cite{varie} and is still a debated problem, here we
will concentrate on the GS properties in the regime of
larger frustration $\alpha \gtrsim 0.6$.
A clear tendency to break the rotational symmetry is suggested by the quasi degeneracy
of the s-wave and d-wave singlets which actually cross each other in the $6\times 6$ cluster.
However, this does not rule out the possible occurrence
of the collinear phase down to $\alpha=0.6$, since the energy gap of the triplet at $(0,\pi)$ 
is shown to decrease with the lattice size. The $(\pi,0)$ singlet gap also decreases with 
size although it remains considerably larger than the lowest singlet gap and in fact 
comparable to the $(\pi,\pi)$ triplet gap which is believed to be finite in the 
thermodynamic limit for  $\alpha > 0.5$. The $(\pi,\pi)$ singlet (not shown in figure) is
much higher in energy. Therefore, from the ordering of the low energy
states we may conclude that: $i)$ rotational symmetry is broken for $\alpha \gtrsim 0.6$;
$ii)$ triplet states are gapped for $\alpha \lesssim 0.62$; $iii)$ the columnar VBC phase
is unlikely to occur, at least for $\alpha \gtrsim 0.62$ and $iv)$ other VBC phases, like a
plaquette state \cite{zhit,sushkov,takano}, are not compatible with the observed ordering 
of levels in the energy spectrum.

The quasi degeneracy between 
the s-wave ($|s>$) and d-wave ($|d>$) singlets in the extended range 
$0.6 < \alpha < 0.7$ allows to consider the two real linear combinations 
of these states as good representations of the symmetry broken phases which are 
physically realized in the thermodynamic limit. This is particularly convenient since 
in a non rotationally invariant state, like $(|s>\pm|d>)/\sqrt{2}$, the rotational
order parameter $\hat O_{\gr{r}}=\hat{\gr{S}}_{\gr{r}} \cdot\hat{\gr{S}}_{\gr{r}+\hat{y}}- 
\hat{\gr{S}}_{\gr{r}}\cdot \hat{\gr{S}}_{\gr{r}+\hat{x}}$, (where $\hat{x}$ and $\hat{y}$ 
are the two primitive vectors of the lattice) may acquire a non-zero value. If 
the two singlets are degenerate and $<\hat O_{\gr{r}}>$ remains finite in the thermodynamic limit, 
rotational symmetry breaking occurs in the model \cite{chi}. 
In the thermodynamic limit, this procedure would be fully equivalent to the usual 
way to evaluate order parameters in terms of the asymptotic behavior 
of correlation functions $<\hat O_{\gr{r}}\hat O_{\gr{0}}>$. However, in small
clusters, we believe that our approach is less affected by finite size effects.
The two (quasi) degenerate states
have vanishing momentum and then the order parameter $<\hat O_{\gr{r}}>$ is translationally invariant
i.e. independent of ${\gr{r}}$.
The numerical results are displayed in Fig. \ref{orderparam} together with the probability to find next
neighbor singlets both on horizontal $P_x$ and vertical $P_y$ bonds.
\begin{figure} \label{fig3}
\includegraphics[width=0.45\textwidth] {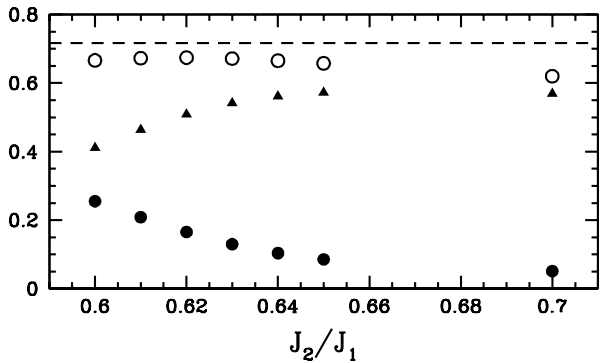}
\vspace{-4mm}
\caption{\label{orderparam}
Properties of the symmetry broken state: Ising order parameter $\hat O$ (triangles); 
Probability to find next neighbors singlets in a given direction; 
$P_\mu=<1-\frac{1}{2}(\hat{\gr{S}}_{\gr{r}}+\hat{\gr{S}}_{\gr{r}+\hat{\mu}})^2>$. 
Empty dots: $P_x$, Full dots: $P_y$. Dashed line: $P_x$ for independent Heisenberg chains.
}
\vspace{-4mm}
\end{figure}

\begin{figure} \label{fig4}
\includegraphics[width=0.48\textwidth, height=0.42\textwidth ]{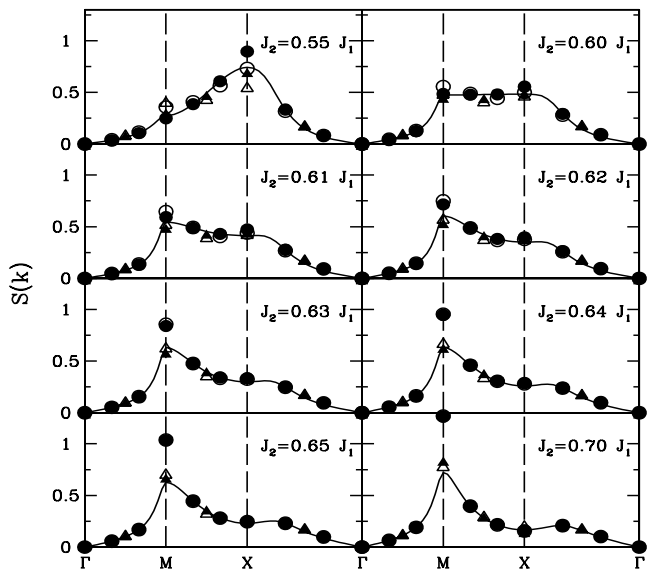}
\vspace{-4mm}
\caption{\label{strut}
Magnetic structure factor along a closed path in the Brillouin zone from 
$\Gamma=(0,0)$ to $M=(0,\pi)$, to $X=(\pi,\pi)$ and back to $\Gamma$. 
Full (empty) triangles: $S(\gr{k})$ evaluated on the lowest s-wave (d-wave) singlet 
for a $4\times 4$ cluster;
Full (empty) dots: $S(\gr{k})$ evaluated on the lowest s-wave (d-wave) singlet 
for a $6\times 6$ cluster 
}
\vspace{-5mm}
\end{figure}
A strong anisotropy is present: The order parameter is large and 
the singlet probability strongly differs in the two spatial directions. 
Remarkably, around $\alpha \sim 0.60$, 
$P_y\sim 0.25$ is compatible with a disordered configuration, while 
$P_x$ gets close to the limit characterizing the Heisenberg chain.
In essence, the whole system seems to behave as a collection 
of spin chains weakly coupled in the transverse direction.
As $\alpha$ grows, $P_x$ keeps almost constant, while $P_y$ decreases,
leaving room to the formation of vertical triplets:
The system is moving toward the collinear phase.

In order to better characterize the phase
diagram we also investigated the behavior of spin correlations. 
The Fourier transform $S(\gr{k})$ of $<S_0^zS_n^z>$
along a closed path in the Brillouin zone has been calculated both for the s-wave 
and d-wave singlet and the results are shown in Fig. \ref{strut}. The close similarity
between the spin correlations in the two states indeed confirms that they both 
contain the same physics. 
At $J_2=0.55\, J_1$ $S(\gr{k})$ exhibits a peak at momentum ($\pi$,$\pi$) 
suggesting that the dominant (short range) correlations are still
antiferromagnetic \cite{CBPS}. 
In the range $0.60 \lesssim \alpha \lesssim 0.62$ $S(\gr{k})$ is remarkably flat 
and does not show significant size dependence. Instead it seems that for larger
frustration $\alpha$ 
the system is going to sustain a transition to a collinear phase. 
Finally, at $\alpha = 0.7$ the $(\pi,0)$ peak in fact grows 
quite substantially with the size, signaling the onset of the collinear order parameter.
A remarkable common feature of the LD results is the collapse of the $4\times 4$ 
and $6\times 6$ 
data on the same smooth curve {\sl except} (possibly) at a 
single wave-vector, which identifies
the dominant periodicity in the spin correlations. 
The LD data then allow for an accurate evaluation of the full momentum 
dependence of the magnetic
structure factor:
we performed a fit of the numerical data (also shown in Fig. \ref{strut})
with a parameterized form inspired by the spin wave theory results \cite{fit}. 
The chosen function represents quite accurately the numerical 
data except at the single wave-vector
where the order parameter sets in and a singular contribution develops in the 
thermodynamic limit.

\section{Conclusions}

In this paper we derived, for the first time, the long wavelength,
low energy effective field theory describing quantum and thermal 
fluctuations in the collinear phase of frustrated $2D$ antiferromagnets. 
The resulting NLSM is written in terms of two fields describing
the local N\'eel order parameter of the two sublattices and 
is invariant under the $O(3)\times Z_2$ symmetry group. 
On the basis of this formalism we are led to predict the
possible occurrence of a non-magnetic ground state breaking
rotational symmetry for suitable values of the frustration ratio.
In order to investigate this possibility, we also performed Lanczos 
diagonalizations. By a careful inspection of the numerical results
we found evidence for the occurrence of the predicted valence bond 
nematic ground state in a region around $\alpha\sim 0.6$. The evaluation
of the magnetic structure factor in small clusters also showed that the 
short range spin correlations of the $J_1-J_2$ model are remarkably
size independent. This observation may be very useful in the interpretation
of accurate neutron scattering data on frustrated $2D$ antiferromagnets. 

According to spin wave theory the $J_1-J_2$ model displays
a first order phase transition at $\alpha _c=0.5$
between a N\'eel and a collinearly ordered region 
in the classical limit ($S \to \infty$).  
When quantum fluctuations are taken into account
an intermediate $SU(2)$ invariant phase is stabilized.
Many different candidates have been proposed as possible ground states 
in this region: gapped 
or gapless spin liquids \cite{figue,CBPS}, VBC's with columnar 
\cite{Poil,singh2,sushkov,kotov}
or plaquette patterns\cite{zhit,sushkov,takano}.

At any finite temperature, the continuous spin rotational symmetry, 
broken in the collinear phase, is restored 
and only the breaking of the symmetry corresponding to the order parameter 
$\sigma= \hat{\gr{n}}_+\cdot\hat{\gr{n}}_-$ 
can in principle survive up to a finite critical temperature defining a
phase transition which lies in the 2D Ising universality class \cite{CCL}.
 The hypothesis that as $T \to 0$ the transition line 
ends in a point different from the 
one corresponding to the onset of the collinear order can not be excluded
 a priori. In such a case, the ground state in a portion of the intermediate 
 $SU(2)$ invariant region may be a valence bond nematic phase with some 
orientational ordering.
This possible scenario is consistent with the suggestions of the analytical results 
based on the NLSM action and has been confirmed by a LD analysis.

Because of the quasi degeneracy in the energy spectrum in a region 
around $\alpha\sim 0.60$, a state characterized by an orientational symmetry breaking 
is very likely to occur, ruling out the fully symmetric spin liquid.
Similarly, the ordering of excited states is not compatible with the plaquette VBC. 
A careful analysis of the low energy spectrum and of the correlation
 functions suggests that 
a zero temperature transition takes place at $\alpha_c\sim 0.62$. 
The transition separates the large $\alpha$ collinear phase and an intermediate
gapped regime breaking the $\pi/2$ rotational symmetry of the lattice. Thus 
on the basis of our investigations we argue that such a phase may be 
conveniently thought of as ``nematic" ordering of valence bonds and anticipates
an isotropic spin liquid (or a VBC) which is likely to occur at lower $\alpha$.
   
A direct transition between the collinear phase and a VBC should be of the first order,
the two phases having different order parameters. Instead the transition to the Ising 
phase may be of second order being related just to the vanishing of the sublattice staggered 
magnetization. Our analysis is not able to discriminate between second order and 
weakly first order transition: Monte Carlo simulations of the NLSM action derived here will be
helpful to supplement LD data in order to clarify this issue.


\begin{thebibliography}{99}
\bibitem{note} This is true within the $O(3)$ NLSM approach where it is assumed that 
the N\'eel field is well defined at all points in (2+1) dimensional space-time.
However if this assumption is relaxed to allow ``hedgehog'' (spatial) singularities 
of the N\'eel field then a dependence on the quantized value of the spin $S$ 
appears and the Berry term may contribute to the Boltzmann weight of the configuration \cite{H1}.
\bibitem{chi} These two conditions together imply that the susceptibility 
associated to the order parameter $\hat O$ diverges.
\bibitem{fit} The fitting function is the square root of the ratio between
two first order polynomials in $\gamma({\gr k})=\cos k_x + \cos k_y$ and
$\gamma'({\gr k})=\cos k_x\cos k_y$.
\end{thebibliography}
\end{document}